\documentclass{elsart}

\usepackage[tbtags]{amsmath}

\usepackage{chicago}

\usepackage{amsfonts}

\usepackage{amssymb}

\usepackage{graphicx}

\newcommand{\tx}[1]{\mathrm{#1}}

\begin{document}

\begin{frontmatter}

\title{Quasispecies can exist under neutral drift at finite population sizes}

\author[CIT]{Robert Forster},
\author[CIT,KGI]{Christoph Adami} and
\author[CIT,KGI,corresp]{Claus O. Wilke}

\address[CIT]{Digital Life Laboratory, California Institute of Technology, Pasadena, California 91125}

\address[KGI]{Keck Graduate Institute of Applied Life Sciences, 535 Watson Drive, Claremont, CA 91711}

\address[corresp]{Corresponding author. Email: wilke@kgi.edu}

\begin{abstract}
  We investigate the evolutionary dynamics of a finite population of RNA
  sequences adapting to a neutral fitness landscape.  Despite the lack of
  differential fitness between viable sequences, we observe typical properties
  of adaptive evolution, such as increase of mean fitness over time and
  punctuated equilibrium transitions.  We discuss the implications of these
  results for understanding evolution at high mutation rates, and extend the
  relevance of the quasispecies concept to finite populations and time scales.
  Our results imply that the quasispecies concept and neutral drift are not
  complementary concepts, and that the relative importance of each is
  determined by a combination of population size and mutation rate.
\end{abstract}


\end{frontmatter}

\section{Introduction}

One of the more interesting aspects of evolution at high mutation rates is the
possible emergence of a {\em quasispecies}.  Originally formulated by Eigen
and Schuster \citeyear{EigenSchuster79}, the quasispecies model describes how
natural selection may act on a group of related genotypes that are coupled via
mutations, rather than on each genotype independently.  This model
is most relevant when the product of population size and genomic mutation rate
exceeds one, so that new mutants are introduced into the population in each
generation. In this context, robustness to mutations can be seen as a
beneficial trait \shortcite{vanNimwegenetal99b}, and selection for this
robustness is invariably associated with the emergence of a quasispecies
\shortcite{Wilke2001b}.  Because RNA viruses have mutation rates in the range
that is relevant for quasispecies theory~\shortcite{Drake93,DrakeHolland99},
quasispecies models have been used to describe the dynamics of RNA virus
populations
\shortcite{Domingo1992,DomingoHolland97,Domingoetal2001,Domingo2002}. However,
this use has generated criticism \shortcite{HolmesMoya2002,Jenkinsetal2001}
because quasispecies theory, as it was originally developed, assumes an
infinite population size and predicts deterministic dynamics. Viral
populations, on the other hand, are finite and subject to stochastic dynamics
and neutral drift.

However, the hallmark of quasispecies dynamics---the existence of a
mutationally coupled population that is the target of selection in its
entirety---does not presuppose an infinite population size or the absence of
neutral drift~\shortcite{vanNimwegenetal99b,Wilke2004}. Rather, infinite
populations were used by Eigen~\citeyear{Eigen1971} and Eigen and
Schuster~\citeyear{EigenSchuster79} to simplify the mathematics of the coupled
differential equations describing the population dynamics. Even though
technically the quasispecies solution of Eigen and Schuster, defined as the
largest eigenvector of a suitable matrix of transition probabilities, only
exists for infinite populations after an infinitely long equilibration period,
it would be wrong to conclude that the cooperative population structure
induced by mutational coupling would disappear when the population is finite.
We show here that quasispecies dynamics are evident in fairly small
populations (effective population size $N_e\leq1000$), and that these dynamics
cross over to pure neutral drift in a continuous manner as the population size
decreases.

We investigate the presence of quasispecies dynamics by simulating populations
of self-replicating RNA sequences, and looking for an unequivocal marker for
quasispecies dynamics in this system, the selection of mutational
robustness~\shortcite{vanNimwegenetal99b,BornbergBauerChan99,Wilke2001b,WilkeAdami2003}.
We choose RNA secondary structure folding \shortcite{Hofackeretal94} as a
fitness determinant because it is a well-studied
model~\shortcite{Huynenetal96,FontanaSchuster98,vanNimwegenetal99b,AncelFontana2000,WilkeAdami2001,Meyersetal2004,Cowperthwaiteetal2005}
in which the mapping from sequence to phenotype is not trivial. The
non-triviality of this mapping is crucial for the formation of a quasispecies,
as we explain in more detail below.  We consider RNA sequences that fold into
a specific target secondary structure as viable, and all other RNA sequences
as non-viable. We choose a neutral fitness model, that is, one where the
replicative speed of all viable sequences is identical and set equal to one,
in order to be able to study quasispecies dynamics exclusively (non-viable
sequences have fitness zero). If there were fitness differences among viable
sequences (i.e., if the fitness landscape contained peaks of different
heights), then adaptive events leading to higher peaks could dominate the
evolutionary dynamics. In the neutral landscape, all peaks are of equal
height, but some peaks are wider than others. Thus, we can describe the RNA
folding landscape as a network of neutral
sequences~\shortcite{Huynenetal96,Reidysetal1997,vanNimwegenetal99b,Reidysetal2001}.
The viable sequences form a network in genotype space, i.e., a graph that
results from including all viable sequences as vertices, and including an edge
between two such vertices if a single mutation can interconvert the two
sequences.  For each sequence, the probability of a mutation being neutral
rather than lethal is called the sequence's {\em neutrality}. In the absence
of differential fitness among the viable sequences, differences in sequences'
neutrality becomes the target of natural selection, that is, we observe
selection for mutational robustness.

In our model, since individual sequences have either fitness 0 or 1, the
average fitness of the population is equal to the fraction of viable sequences
in the population. An increase in the average fitness therefore indicates that
the sequences in the population have become more robust to mutations
\shortcite{vanNimwegenetal99b}. In order to detect selection for mutational
robustness, we look for abrupt transitions of the adapting population to
higher average fitness, after allowing for an initial equilibration period.
For a purely neutrally drifting swarm of sequences such transitions cannot
exist, although stochastic effects can mimic such transitions if the
population size is small.  If we observe transitions in average fitness above
the background level expected from stochastic fluctuations, we can ascribe
these transitions to the discovery of a region of higher neutrality on the
neutral network.  In this case, the transition to higher average fitness
corresponds to the outcompetition of the previous quasispecies by a more
mutationally robust one.

\section{Materials and Methods} 

We consider a population of fixed size $N$ composed of asexual replicators
whose probability of reproduction in each generation is proportional to their
fitness (Wright-Fisher sampling).  The members of the population are RNA
sequences of length $L=75$, and their fitness $w$ is solely a function of
their secondary structure.  Those that fold into a specific target secondary
structure are deemed viable with fitness $w=1$, while those that fold into any
other shape are non-viable ($w=0$).  The average fitness $\langle w\rangle$
of the population is therefore the fraction of living members out of the total
population.  RNA sequences are folded into the minimum free energy structure
using the Vienna Package~\shortcite{Hofackeretal94}, and dangling ends are
given zero free energy \shortcite{Walteretal94}.  For a given simulation, an
initial RNA sequence is selected uniformly at random and its minimum-energy
secondary structure defines the target structure for this simulation, thereby
determining a neutral network on which the population evolves for a time of
$T=50$,000 generations.  Mutations occur during reproduction with a fixed
probability $\mu$ per site, corresponding to an average genomic mutation rate
$U=\mu L$. 

Our simulations spanned a range of genomic mutation rates and population
sizes, and we performed $50$ independent replicates for each of the pairs
$(U,N)$, starting each with a different randomly chosen initial sequence.  To
study mutation rate effects, we considered a fixed population size of
$N=1000$, across a range of genomic mutation rates, using $U=0.1, 0.3, 0.5,
1.0,$ and $3.0$.  To study effects due to finite population size, we
considered a fixed mutation rate of $U=1.0$, using population sizes of $N=30,
100, 300,$ and $1000$.

The neutrality of a sequence was determined by calculating the fraction of
mutations that did not change the minimum-energy secondary structure. Thus, if
$N_\nu$ of all $3L$ one-point mutants of a sequence retain their structure,
the neutrality of that sequence is given by $\nu=N_{\nu}/3L$.  Because
sequences that don't fold into the target structure have zero fitness, a
sequence's neutrality is equal to the mean fitness of all possible single
mutants.  We recorded the population's average fitness every generation, while
the population's average neutrality, being much more computationally
expensive, was calculated only at the start and end of each replicate. For
illustrative purposes, select replicates of interest were recreated using the
original random seed, and the population's neutrality was recorded every 100
generations.

To observe the signature of natural selection acting within our system, we
derive a statistical approach to identify transitions in the population's
average fitness $\langle w\rangle$.  If a beneficial mutation appears and is
subsequently fixated in the population, we expect to observe a step increase
in the population's average fitness. We emphasize again that such selective
sweeps must be due to periodic selection of quasispecies for increased
mutational robustness, since there are no fitness differences between
individual genotypes.

In light of the fluctuations in the population's average fitness due to
mutations and finite population effects, we employ statistical methods to
estimate the time at which the beneficial mutation occurred and associate a
$p$-value with our level of confidence that a transition has occurred.  Our
approach can be thought of as a generalization of the test for differing means
between two populations (those before and after the mutation), except that the
time of the mutation's occurrence is unknown {\it a priori}.  For a full
derivation and discussion of our approach, see the Appendix. While our
alogrithm can be applied recursively to test for and identify multiple
transitions that may occur in a single simulation, unless otherwise noted, we
considered only the single most significant step found.

\section{Results} 

Because replicates were initialized with $N$ (possibly mutated) offspring of
the randomly chosen ancestor, the simulation runs did not start in
mutation--selection balance. Typically, we observed an initial equilibration
period of 50 to 200 generations, after which the population's fitness and
neutrality stabilized, with fluctuations continuing with magnitude in
proportion to the mutation rate.  As predicted by van Nimwegen et
al.~\citeyear{vanNimwegenetal99b}, during the equilibration period, we
observed in most replicates beneficial mutations that increased the
equilibrium level of both average fitness and neutrality. (Throughout this
paper, by beneficial mutations we mean mutations that increase a sequence's
neutrality, and thus indirectly the mean fitness of the population. There are
no mutations that increase the fitness of a viable sequence beyond the value 1
in our system.) These mutations led to the initial formation of a quasispecies
on a high-neutrality region of the neutral network. For the remainder of this
paper, we are not interested in this initial equilibration, but in transitions
towards more densely connected areas of the neutral network once the initial
equilibration has occurred.

\begin{figure*}[thb]
\centerline{\includegraphics[width=5in]{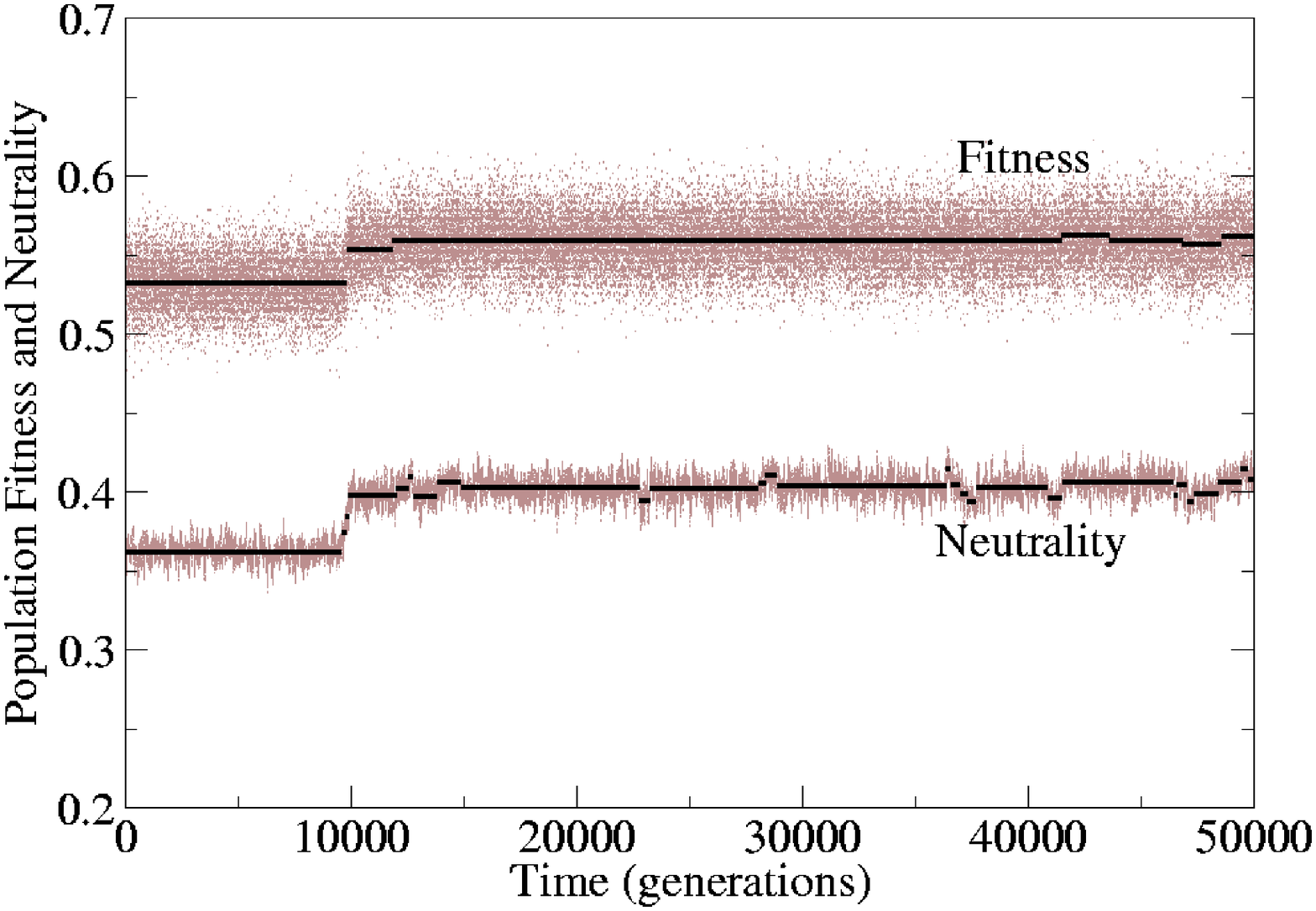}} 
\caption{\label{fig:step} Average fitness and neutrality of a population
  during a single simulation at a genomic mutation rate of $U=1.0$. At
  $t=9814$, a $5\%$ increase in the population's average fitness occurs at the
  $p<10^{-7}$ level, with a corresponding transition in the population's
  average neutrality. Smaller transitions occur throughout the simulation run.
  The solid lines indicate the epochs of constant fitness and neutrality, as
  determined by our step-finding algorithm. As explained in the Appendix, the
  application of this algorithm to the neutrality data is for illustrative
  purposes only. Because of temporal autocorrelations in the neutrality, not
  all steps that the algorithm identifies are statistically significant.}
\end{figure*}

To determine if such a transition has occurred, we need a method to
distinguish significant changes in the population's mean fitness from apparent
transitions caused by statistical fluctuations.  We devised a statistical test
(see Appendix for details) that can identify such transitions and assign a
$p$-value to each event.  We found that transitions to higher average fitness
occurred in over 80\% of simulations across all mutation rates studied, if we
considered all transitions with $p$-values of $p < 0.05$.
Figure~\ref{fig:step} shows a particularly striking example of such a
transition ($p$-value $\le 10^{-7}$), where a 5.0\% increase in average
fitness occurs at $t=9814$.  A similar analysis of the average neutrality (not
usually available, but computed every generation specifically in this case)
finds an increase of 11.2\% occurring at $t=9876$, with the same level of
confidence.  The multiple transitions shown in the Figure~\ref{fig:step} are
the results of recursively applying our step-finding algorithm until no steps
are found with $p<0.05$.

Depending on the mutation rate, a step size as little as 0.04\% in the
population's average fitness could be statistically resolved in a background
of fitness fluctuations several times this size.  For comparison, typical
noise levels, as indicated by the ratio of the standard deviation of the
fitness to its mean, ranged from 0.7 to 6.6\% over the mutation rates studied.
Note that fluctuations in the neutrality level are much smaller, due to the
additional averaging involved. However, because neutrality is much more
expensive computationally, and would also be difficult to measure in
experimental viral populations, we used mean fitness as an indicator of
transitions throughout this paper.

\begin{figure*}[thb]
\centerline{\includegraphics[width=5in]{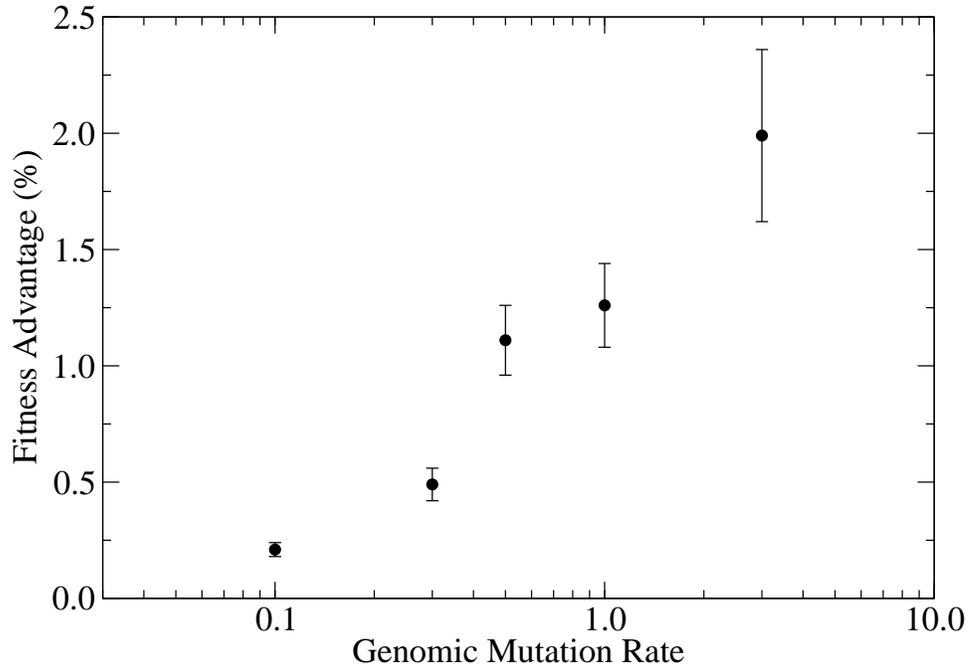}}

\caption{\label{fig:avg_step} Average step size as a function of genomic
  mutation rate ($U=0.1,0.3,0.5,1.0,3.0$).  Step size is measured by percent
  increase in the population's fitness, with only runs significant at the
  $p<0.05$ level shown.  Error bars are standard error.}
\end{figure*}

Figure~\ref{fig:avg_step} shows the average size of the most significant step
observed as a function of the mutation rate.  At low mutation rates, such as
$U=0.1$, the smaller observed step size corresponds to the fact that $90\%$ of
the population is reproducing without error, and hence improvements in
neutrality can only increase the population's fitness in the small fraction of
cases when a mutation occurs. At higher mutation rates the step sizes
increase, reflecting the larger beneficial effect of increased neutrality
under these conditions.

\begin{figure*}[htb]
\centerline{\includegraphics[width=5in]{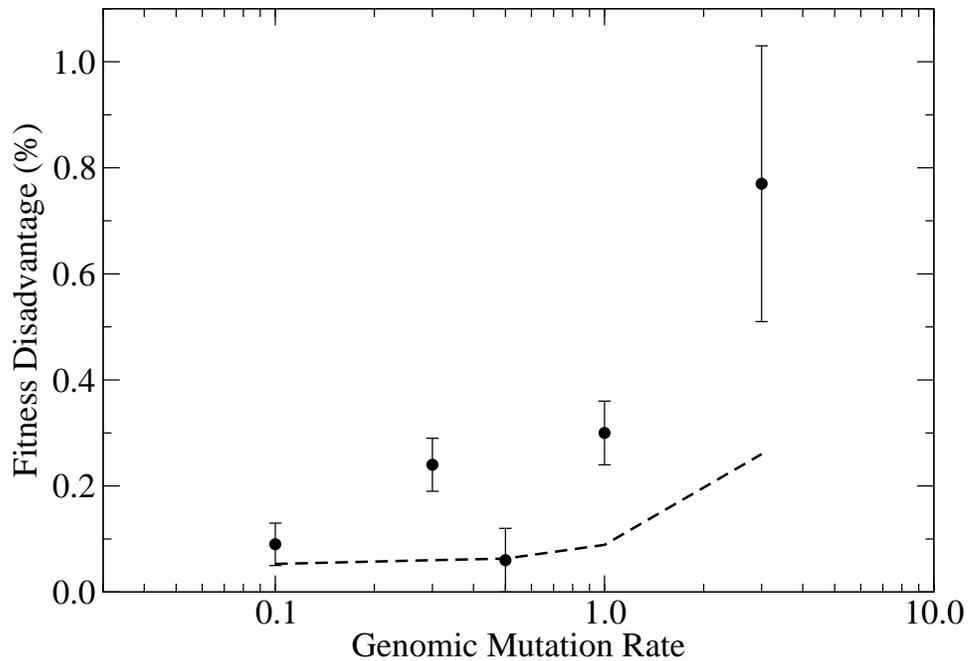}}

\caption{\label{fig:neg_step} Average step size $|s|$ of statistically
  significant drops in fitness (at the $p<0.05$ level).  Step size is measured
  by relative decrease in population fitness, and error bars are standard
  error.  The dotted line indicates $2|s|=1/N_e$, a selective
  disadvantage consistent with neutral drift in a finite population.
  $N_e$ is the average number of living members of the population
  (effective population size).}
\end{figure*}

In about $10\%$ of all simulations with statistically significant changes in
fitness, the most significant change in fitness was actually a step {\it
  down}, that is, a fitness loss, rather than the increase in fitness
typically observed. Negative steps in average fitness occur due to stochastic
fixation of detrimental mutations at small
population sizes \cite{kimura62}. These negative
fitness steps, however, are generally much smaller than the typical positive
step size. The average size of these
negative steps was between 0.09 and 0.77\%, compared with an average positive
step size between 0.27 and 2.33\% (see Figure~\ref{fig:neg_step}).

\begin{figure*}[htb]
\centerline{\includegraphics[width=5in]{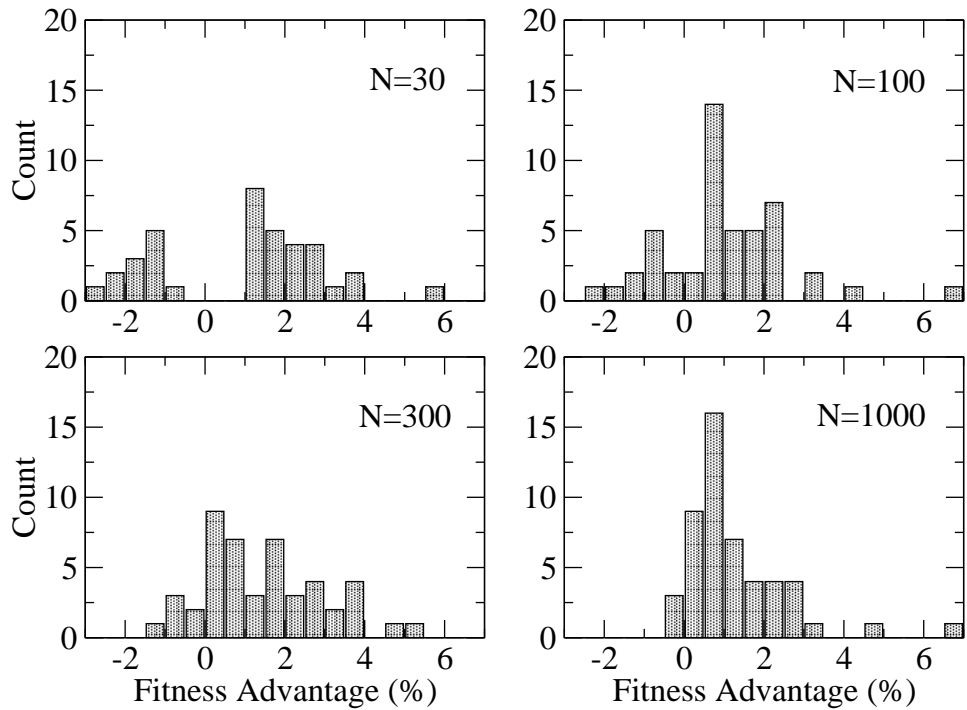}} 

\caption{\label{fig:step_dist} Distribution of sizes of the most significant
  step (at $p<0.05$) in each run, out of 50 runs at four population sizes
  ($U=1$).  At small population sizes, the distribution is almost symmetric
  about zero since most mutations are of less benefit than the $1/N_e$
  probability of fixation due to drift.  At large sizes, selection is evident
  from the positively skewed distribution.}
\end{figure*}

\begin{figure*}[htb]
\centerline{\includegraphics[width=5in]{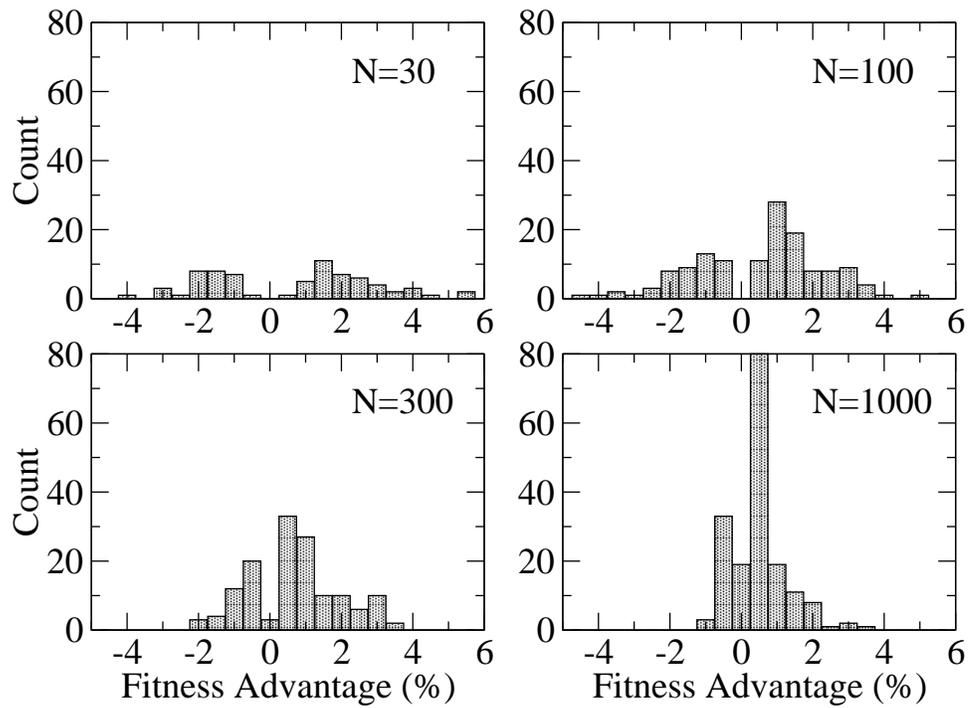}} 

\caption{\label{fig:step_dist_all} Distribution of sizes of all significant
  steps (at $p<0.05$) in each run, out of 50 runs at four population sizes
  ($U=1$).  While these distributions are more symmetrical than those of
  Fig.~\ref{fig:step_dist}, a substantial skew towards positive step sizes is
  still evident for the larger population sizes.}
\end{figure*}

We specifically studied the role of finite population size and its effects on
neutral drift by considering populations of size $N=30$, 100, 300, and 1000 at
a constant genomic mutation rate of $U=1.0$.  We again performed 50 replicates
at each population size, and the distribution of statistically significant
step sizes are shown in Fig.~\ref{fig:step_dist} (biggest step only)
and Fig.~\ref{fig:step_dist_all} (all steps).  While the larger population's
distributions show a clear bias towards positive steps in fitness, the
distributions become increasingly symmetric about zero for smaller population
sizes.  A gap around zero fitness change becomes increasingly pronounced in
smaller populations, as the fluctuations in fitness due to finite population
size preclude us from statistically distinguishing small step sizes from the
null hypothesis that no step has occurred.

\begin{figure*}[htb]
\centerline{\includegraphics[width=5in]{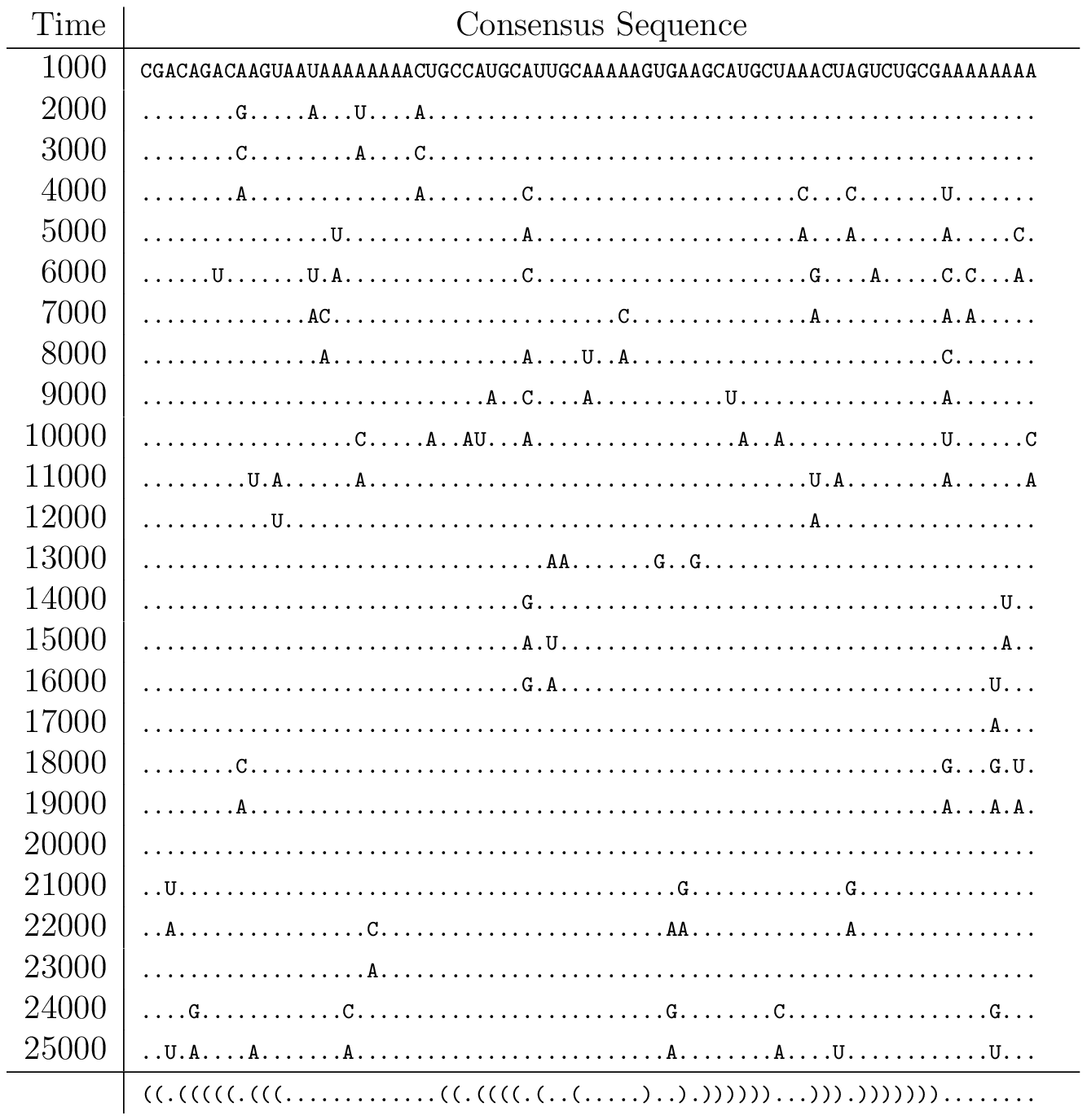}} 

\caption{\label{fig:consensus}Change in the consensus sequence over time, from
  the same simulation run as presented in Fig.~\ref{fig:step}. Dots in the
  alignment indicate that the base at this position is unchanged from the
  previous line. The bottom row shows the target secondary structure in
  parentheses notation for reference.}
\end{figure*}

We also kept track of the consensus sequence in our simulations, to determine
whether the population underwent drift while under selection for mutational
robustness.  In the runs with $N=1000$, the consensus sequence accumulated on
average one substitution every 2 to 3 generations. As such rapid change might
be caused by sampling effects, we also studied the speed at which the
consensus sequence changed over larger time windows.  Using this method with
window lengths of 50 and 100 generations, we found that the consensus sequence
accumulated one substitution every 10 to 20 generations (window size 50
generations) or 15 to 30 generations (window size 100 generations).  Thus we
find that the populations continue to drift rapidly throughout the simulation
runs, and never settle down to a stable consensus sequence.
Figure~\ref{fig:consensus} shows the evolution of the consensus sequence over
time for the same simulation run as shown in Fig.~\ref{fig:step}.

Finally, to confirm that our finite population was not sampling the entire
neutral network during our simulations, we estimated the average size of the
neutral network.  We can represent each RNA secondary structure in
dot-and-parenthesis notation, where matched parentheses indicate a bond
between the bases at those points in the sequence and dots represent unpaired
bases.  The number of valid strings of length $L$ can be counted using Catalan
numbers $\tx{Cat}(n) =\binom{2n}{n}/(n+1)$, which give the number of ways to
open and close $n$ pairs of parentheses \cite{vanLintWilson2001}.  Since there
are $4^L$ possible RNA sequences, we obtain for the average network
\begin{equation}\label{eq:network-size}
\langle \tx{network \ size}\rangle = 4^L\Big/\sum\limits_{i=0}^{[L/2]} \tx{Cat}(i) \binom{L}{L-2i} \approx 1.1 \times 10^{12}
\end{equation}
for $L=75$. This expression is a lower bound to the true average network size,
because the denominator counts some unphysical structures, such as hairpins
wiht fewer than 3 bases.  For comparison, the number of possible distinct
genotypes that can appear in each simulation is maximally $N T = 5 \times 10^7
$.

\section{Discussion}

In the study of varying mutation rates, the observed increases in the
population's fitness in almost all replicates demonstrate the action of
natural selection.  Since all viable sequences are neutral and hence enjoy no
reproductive fitness advantage, this selection acts on increasing the
population's robustness to mutations through increases in its average
neutrality (as seen in Figure~\ref{fig:step}).  Thus, these results show
evidence that a quasispecies is present in almost all cases, even though the
difference between a randomly drifting swarm and a population structured as a
quasispecies decreases as the population size and mutation rate decrease. 
Our results also show evidence of neutral drift leading to the fixation of
detrimental mutations in some populations.  The negative steps observed
(Figure~\ref{fig:neg_step}) were comparable in size to $1/N_e$, the
probability of a neutral mutation drifting to fixation.

In the study of varying population sizes, the distribution of mutational
effects on fitness showed an increasing bias towards beneficial rather than
detrimental mutations as the population's size increased
(Figures~\ref{fig:step_dist},~\ref{fig:step_dist_all}).  At population sizes
100, 300, and 1000, the clear positive bias of mutational effects illustrates
the presence of a quasispecies, where natural selection is able to act to
improve the population's neutrality and hence its robustness to mutations.  As
the fluctuations in fitness due to small population size become more
significant, selection for neutrality becomes less relevant when the $1/N_e$
sampling noise exceeds the typical step size of 1\%.  At the smallest
population size of $30$, there still seems to be a bias towards beneficial
mutations, but the evidence is less clear and more replicates are probably
necessary to observe a clear signal of quasispecies dynamics.

Since the average network size is many orders of magnitude larger than the
number of sequences produced during a simulation, we know that the system is
non-ergodic and the population cannot possibly have explored the whole neutral
network.  Moreover, Reidys et al.~\citeyear{Reidysetal1997} studied the
distribution of neutral network sizes in RNA secondary structure and found 
that they obey a power law distribution, implying that there are a small
number of very large networks, and many smaller networks. As a consequence,
choosing an arbitrary initial sequence will more likely result in the choice
of a large network. Therefore, Eq.~\eqref{eq:network-size} is effectively a
lower bound on the sizes of the networks we actually sampled.

We have shown that quasispecies dynamics is not confined to the
infinite population-size limit. Instead, one of the hallmarks of quasispecies
evolution---the periodic selection of more mutationally robust quasispecies in
a neutral fitness landscape---occurs at population sizes very significantly
smaller than the size of the neutral network they inhabit. Despite small
population sizes, if the mutation rate is sufficiently high (in the
simulations reported here, it appears that $N U\gtrsim30$ is sufficient),
stable frequency distributions significantly different from random develop on
the partially occupied network in response to mutational pressure. Most
importantly, we have shown that genetic drift can occur simultaneously with
quasispecies selection, and becomes dominant as $N U$ decreases. Thus, the
notion that genetic drift and quasispecies dynamics are mutually exclusive
cannot be maintained. Instead, we find that both quasispecies dynamics and
neutral drift occur at all finite population sizes and mutation rates, but
that their relative importance changes.

The existence of a stable consensus sequence in the presence of high sequence
heterogeneity has long been used as an indicator of quasispecies dynamics
\shortcite{Domingoetal78,Steinhaueretal89,Eigen96,Jenkinsetal2001,Domingo2002}.
Here, we have shown that quasispecies dynamics can be present while the
consensus sequence changes over time. In our simulations, the consensus
sequence drifts randomly, in a manner uncorrelated with the transitions in
average fitness that we detect.  Thus, quasispecies dynamics does not require
individual mutants to be stably represented in the population, nor does it
require a stable consensus sequence.

The population structure on the neutral network is strongly influenced by the
mutational coupling of the genotypes that constitute the quasispecies. This
coupling arises because mutations are not independent in the landscape we
studied. Rather, as in most complex fitness landscapes, single mutations at
one locus can affect the fitness effect of mutations at another (a sign of
epistasis,~\shortciteNP{Epistasis2000}). In the neutral fitness landscape
investigated here, mutations at neutral or non-neutral (i.e., lethal) sites
can influence the neutrality of the sequence. The absence of epistatic
interactions between the neutral mutations in the fitness landscape studied by
Jenkins et al.~\citeyear{Jenkinsetal2001} implies the absence of quasispecies
dynamics in these simulations. Theoretical arguments show that a
non-interacting neutral region in a genome does not alter the eigenvectors of
the matrix of transition probabilities, and therefore cannot affect
quasispecies dynamics.

Using fitness transitions in neutral fitness landscapes as a tool to diagnose
the presence of a quasispecies has a number of interesting consequences from a
methodological point of view. Clearly, because selection for robustness is a
sufficient criterion for quasispecies dynamics but not a necessary one, the
absence of a transition does not imply the absence of a quasispecies. At the
same time, as the population size decreases, fluctuations in fitness become
more pronounced, rendering the detection of a transition more and more
difficult. Theoretical and numerical arguments suggest that small populations
at high mutation rate cannot maintain a
quasispecies~\shortcite{vanNimwegenetal99b,Wilke2001a}, so the disappearance
of the mutational robustness signal at small population sizes is consistent
with the disappearance of the quasispecies. However, the type of analysis
carried out in this work does not lend itself to detecting quasispecies in
real evolving RNA populations, because the fitness landscape there cannot be
expected to be strictly neutral. Instead, transitions from one peak to another
of different height~\cite{BurchChao2000,Novella2004} are likely to dominate.
Quasispecies selection transitions such as the one depicted in
Fig.~\ref{fig:step} can, in principle, be distinguished from peak-shift
transitions in that every sequence before and after the transition should have
the same fitness. Unfortunately, pure neutrality transitions are likely to be
rare among the adaptations that viruses undergo, and the data necessary to
unambiguously identify them would be tedious if not impossible to obtain.

Our simulations provide evidence of selection for mutational robustness
occurring in the form of increased neutrality of RNA sequences for population
sizes far below the size of the neutral network that the sequences inhabit.
Such increased neutrality was recently found in a study that compared evolved
RNA sequences to those deposited in an aptamer
database~\shortcite{Meyersetal2004}. For example, the comparison showed that
human tRNA sequences were significantly more neutral, and hence more robust to
mutations, than comparable random sequences that had not undergone
evolutionary selection. However, we must caution that while in our
simulations, selection for mutational robustness is the only force that can
cause the sequences to become more mutationally robust, in real organisms
other forces, for example selection for increased thermodynamic stability
\shortcite{Bloometal2005}, could have similar effects.

An experimental system that is quite similar to our simulations, probably more
so than typical RNA viruses, is that of {\em viroids}---unencapsidated RNA
sequences of only around $300$ bases---capable of infecting plant hosts.
Viroid evolution appears to be limited by the need to maintain certain
secondary structural aspects~\cite{KeeseSymons1985}, which is consistent with
our fitness assumptions.  Furthermore, in Potato spindal tuber virus (PSTVd),
a wide range of single and double mutants are observed to appear after a
single passage~\cite{OwensThompson2005}, suggesting that a quasispecies
rapidly forms under natural conditions.  Viroids may have agricultural
applications as they are capable of inducing (desirable) dwarfism in certain
plant species~\shortcite{Huttonetal2000}, and as such, a better understanding
of their evolutionary processes may help to direct future research efforts.

Making the case for or against quasispecies dynamics in realistic, evolving
populations of RNA viruses, or even just self-replicating RNA molecules, is
not going to be easy. As the presence of an error threshold
(\citeNP{WagnerKrall93,Wiehe97}; see also discussion in \citeNP{Wilke2005})
or the persistence of a consensus sequence (this work) have been ruled out as
a diagnostic, we have to look for markers that are both unambiguous and easy
to obtain. Selection for robustness may eventually be observed in natural
populations of adapting RNA viruses or viroids, but up to now, no such signals
have been reported. Thus, while we can be confident that small population
sizes do not preclude quasispecies dynamics in RNA virus populations, on the
basis of current experimental evidence, we cannot decide whether quasispecies
selection takes place in RNA viruses.

\section{Conclusions}

Quasispecies effects are not confined to deterministic systems with infinite
population size, but are readily observed in finite---even small---populations
undergoing genetic drift. We find a continuous transition from very small
populations, whose dynamics is dominated by drift, to larger populations,
whose dynamics is dominated by quasispecies effects. The crucial parameter is
the product of effective population size and genomic mutation rate, which
needs to be significantly larger than one for quasispecies selection to
operate. However, experimental evidence for these theoretical findings is
currently not available, and will most likely be hard to obtain, because the
differences in the dynamics of populations that are simply drifting and
populations that are under quasispecies selection can be quite subtle. Thus, a
dedicated experimental effort is needed to demonstrate quasispecies selection
in natural systems.

\section{Acknowledgments}

This work was supported in part by NIH grant AI 065960.


\newpage

\begin{appendix}

\section{Appendix}

\subsection{The distribution of the population's average fitness as a random variable}

In equilibrium, the distribution of the population's average fitness follows
from Wright-Fisher sampling. Define $\pi$ as the probability that a sequence's
offspring will be viable.  Without resorting to an explicit form for $\pi$,
equilibrium and a uniform mutation rate imply that all sequences reproduce
successfully with the same probability $\pi$ (which is in general a function
of the mutation rate and the mean neutrality of the population). Denote
further the expected value of a random variable $x$ as $E[x]$ and its variance
as $V[x]$.  If we take fitness $w_i$ of the $i$th offspring in our population
as a random variable, the neutral fitness landscape implies that $w_i$ takes
only values $0$ or $1$, where $w_i=1$ occurs with probability $\pi$.  The
distribution of $w_i$ is therefore a Bernoulli distribution with probability
of success $\pi$, and we have $E[w_i]=\pi$ and $V[w_i]=\pi(1-\pi).$

We now consider the average fitness of the population in equilibrium,
$\langle w\rangle$, defined as $\langle w\rangle=\frac{1}{N}\sum_{i=1}^{N}w_i.$
By the Central Limit Theorem, the distribution of $\langle w\rangle$ will
approach a normal distribution $N[\mu, \sigma^2]$ as
$N\rightarrow \infty$, and this limit will be reached well before $N=1000$
(typically $N\pi, N(1-\pi)>5$ is sufficient \cite{Rice1994}, and
this condition is easily satisfied under all conditions studied). Thus, 
$\langle w\rangle$ follows a normal distribution with mean $\mu=E[w]=\pi$ and
variance $\sigma^2=V[w]=\pi(1-\pi)/N$.

\begin{figure*}[htb]
\centerline{\includegraphics[width=5in]{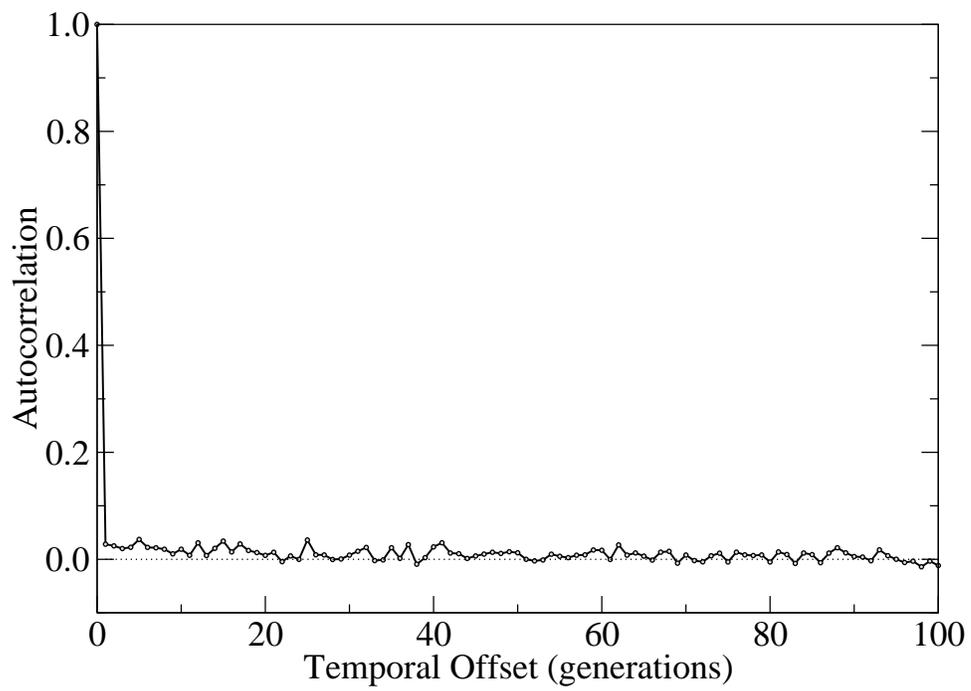}}

\caption{\label{fig:autocorr}Temporal autocorrelation function for the first
  equilibrium period shown in Figure~\ref{fig:step} ($t=200-9814$).}
\end{figure*}

To confirm these assumptions hold, we computed the fitness autocorrelation
function within a period of equilibrium.  Figure~\ref{fig:autocorr} shows
the autocorrelation function for the first equilibrium period shown in
Figure~\ref{fig:step} ($t=200-9814$).  The autocorrelation drops almost
immediately to a mean of nearly zero, and has a noise level $\sigma \approx
1-2\%$, consistent with the variation of $w$ over the time period in question.
Similar results hold for each period of fitness equilibrium shown in
Figure~\ref{fig:step}.  In contrast, the population's average neutrality
showed significant autocorrelations.  While we included the neutrality
transitions in Figure~\ref{fig:step} for illustrative purposes, this lack of
independence suggests that not all the neutrality steps identified are
statistically significant.

\subsection{Identifying jumps in average fitness}

Motivated by our observations, we seek to characterize the rapid transitions
of the population from lower to higher neutrality states.  We derive a
statistical test for identifying such transitions {\it a priori} in time
series data, and associating a $p$-value to measure the confidence level of
such a transition occurring by chance.

Consider a time series $w(t) \sim N[\mu, \sigma^2]$ measured at $T$ sequential
points in time.  To test the hypothesis of equal means between two specified
time periods $[1,n]$ and $[n +1, T]$ is straightforward, and we will assume
equal variances for simplicity.  We consider the average value of $w$ over the
two periods separately, and consider the sample means $Y_i$ over the two
different time periods, defined by $Y_1 = \frac{1}{n} \sum_{t=1}^{n} w(t)$ and
$Y_2 = \frac{1}{T-n} \sum_{t=n+1}^{T} w(t)$. These sample means will be
normally distributed, $Y_1 \sim N[\mu, \frac{\sigma^2}{n}]$ and $Y_2 \sim
N[\mu, \frac{\sigma^2}{T-n}]$. Our null hypothesis is that the means will be
equal between the two periods. To test this null hypothesis, we consider the
difference between the sample means $D=Y_2-Y_1$, and ask whether the observed
difference can be explained merely by chance, that is, whether the
distribution of $D$ is consistent with $D\sim N[0, \sigma_D^2]$. Here,
$\sigma^2_D$ is the sum of the variances of $Y_1$ and $Y_2$, that is,
$\sigma^2_D=\sigma^2T/[n(T-n)]$.

Thus, under the null hypothesis, the difference of observed means $D$ is
normal with zero mean and known variance, and the associated $p$-value can
be obtained by looking up the probability of $Z=D/\sigma_D$ exceeding its
observed value in a cumulative distribution table.

We now consider the case of finding the most significant breakpoint in the
time series $[1,T]$ when the division into two periods is unspecified.
Letting $n$ parameterize the number of data points in the first interval, we
can consider the above analysis as a function of $n$.  The highest
significance is attained by choosing the maximum value of $D(n)/\sigma_D(n)$,
where the difference of means and its variance must be calculated for all $n$
in $[1,T-1]$.  Let $p_n$ represent the $p$-value associated with this maximum
$n$. We wish to know the probability that this maximum level of
significance will occur merely by chance due to the fluctuations in $w(t)$.
Given $T-1$ independent trials with probability $p_n$ of exceeding our maximum
level of significance, we see that the probability of all of these trials
resulting in a smaller significance than that of $p_n$ is 
\begin{align}
 \tx{Pr}[\text{all $T-1$ of the $p_i$ satisfy $p_i < p_n$}] &= (1-p_n)^{T-1} \notag\\
& \approx 1-(T-1)p_n \quad \text{for} \;\,p_n\ll 1.
\end{align}
From this probability, we calculate the $p$-value associated with any $p_i$
exceeding our $p_n$ by chance alone, using the above probability:
\begin{align}
 p &= \tx{Pr}[\text{at least one $p_i$ has $p_i>p_n$}] \notag\\
   &= 1 - \tx{Pr}[\text{all $T-1$ of the $p_i$ satisfy $p_i < p_n$}]\notag\\
   &= 1- (1-p_n)^{T-1} \approx (T-1)p_n\,.
\end{align}
Note that the $T-1$ other choices of breakpoints are by no means independent
of each other, as they all refer to the same underlying fitness data, $w(t)$.
These correlations reduce the number of effective degrees of freedom, and
hence the $T-1$ factor will be a conservative overestimate of the actual
$p$-value.  If multiple transitions are expected, this algorithm can be
repeated on each subinterval to determine whether further breakpoints are
consistent with the given level of statistical confidence.

\end{appendix}

\newpage

\end{document}